\documentclass[lettersize,journal]{IEEEtran}
\usepackage{amsmath,amsfonts}
\usepackage{algorithmic}
\usepackage{algorithm}
\usepackage{array}
\usepackage[caption=false,font=normalsize,labelfont=sf,textfont=sf]{subfig}
\usepackage{textcomp}
\usepackage{stfloats}
\usepackage{url}
\usepackage{verbatim}
\usepackage{graphicx}
\usepackage{multirow}
\usepackage{multicol}
\usepackage{cleveref}

\usepackage[inline]{enumitem}
\usepackage{xcolor}
\usepackage{xcolor,colortbl}
\usepackage[
  separate-uncertainty = true,
  multi-part-units = repeat,
  parse-numbers=false
]{siunitx}

\usepackage{amsthm}
\usepackage{empheq} 
\usepackage{booktabs}
\usepackage{pifont}
\usepackage{amssymb}
\newcommand{\xmark}{\text{\ding{55}}}

\newtheorem{assumption}{Assumption}

\newtheorem{definition}{Definition}

\usepackage[style=ieee, hyperref=false]{biblatex}
\AtBeginBibliography{\footnotesize}

\AtEveryBibitem{
    \clearfield{url}
    \clearfield{urlyear}
    \clearfield{urlmonth}
    \clearfield{doi}
    \clearfield{issn}
}
\begin{document}

\title{Automated Lane Merging via Game Theory and Branch Model Predictive Control}

\author{Luyao Zhang$^{1}$, Shaohang Han$^{2}$ and Sergio Grammatico$^{1}$
\thanks{*This work is partially supported by NWO under project AMADeuS.}
\thanks{Luyao Zhang and Sergio Grammatico are with the Delft Center for Systems and Control, TU Delft, The Netherlands.
        {\tt\small \{l.zhang-7, s.grammatico\}@tudelft.nl}
        }%
\thanks{$^{2}$Shaohang Han is with the
Division of Robotics, Perception and Learning, KTH
Royal Institute of Technology, Sweden. {\tt\small shaohang@kth.se}.}%
}



\maketitle

\begin{abstract}
We propose an integrated behavior and motion planning framework for the lane-merging problem.
The behavior planner combines search-based planning with game theory to model vehicle interactions and plan multi-vehicle trajectories. 
Inspired by human drivers, we model the lane-merging problem as a gap selection process and determine the appropriate gap by solving a matrix game.
Moreover, we introduce a branch model predictive control (BMPC) framework to account for the uncertain equilibrium strategies adopted by the surrounding vehicles, including Nash and Stackelberg strategies.
A tailored numerical solver is developed to enhance computational efficiency by exploiting the tree structure inherent in BMPC.
Finally, we validate our proposed integrated planner using real traffic data and demonstrate its effectiveness in handling interactions in dense traffic scenarios. 
The code is publicly available at: \url{https://github.com/SailorBrandon/GT-BMPC}.
\end{abstract}


\section{Introduction} \label{sec:intro}
\subsection{The lane-merging problem}
\IEEEPARstart{T}{he} last two decades have seen prosperous progress in autonomous driving technology.
While automated vehicles have demonstrated the ability to operate in various daily traffic scenarios, they still face challenges when navigating through highly interactive environments.
This is primarily because automated vehicles often struggle to account for the diverse behaviors exhibited by the surrounding vehicles.
In this paper, we focus on the lane-merging scenario, where the automated vehicle needs to promptly find a suitable gap and understand whether the surrounding vehicles are willing to yield or not.
In traditional methods, the motion predictor provides the predicted trajectory of a surrounding vehicle, indicating its driving intention, e.g., yielding. Based on these predicted trajectories, the motion planner generates a safe and comfortable trajectory for the ego vehicle \cite{dingSafeTrajectoryGeneration2019, fan_baidu_2018}.
Since motion prediction and planning are decoupled in this hierarchical structure, it often overlooks the mutual interactions between the ego vehicle and its surroundings. 
Consequently, the generated motion plans can be overly conservative.
Although recently developed learning-based approaches \cite{zengEndToEndInterpretableNeural2019, chenLearningAllVehicles2022} consider such interactions, 
they inherently come with issues such as distribution shifts \cite{thagaard2020can,filos2020can} and a lack of interpretability, potentially leading to a loss of safety guarantees.
In this paper, we aim to develop a model-based interaction-aware planning system that integrates motion planning with motion prediction to handle the uncertain behaviors of the surrounding vehicles.

Interactive environments present two primary challenges: 
\begin{enumerate*}[label=(\roman*)]
\item accurately modeling vehicle interactions, and \item effectively handling the diverse behavior modes of surrounding vehicles.
\end{enumerate*}
To address the first challenge, researchers have proposed a game-theoretic planning framework to capture the mutual influence among multiple players.
Each player in a game has its individual objective function, which depends not only on its own action but also on the actions of other players, and the goal of each player is to optimize its objective function.
Previous research has extensively explored equilibrium solutions for autonomous driving. Some studies focus on jointly planning trajectories for all vehicles by seeking a Nash equilibrium \cite{FridovichILQG2020, cleachALGAMESFastSolver2020}. 
Other approaches utilize semantic-level actions as strategies to leverage domain knowledge of the lane-merging problem.
Among them, some studies investigate the Stackelberg equilibrium with a leader-follower game structure \cite{ liuInteractionAwareTrajectoryPrediction2023, weiGameTheoreticMerging2022}.
In contrast to the leader-follower structure, a Nash game treats all agents equally \cite{cleachALGAMESFastSolver2020}.
Several methods based on Nash games have been proposed in \cite{lopezGameTheoreticLaneChangingDecision2022, HangHumanLike2021, CoskunRHMarkovGame2019}, assuming that all players adhere to one specific equilibrium. 
In practice, however, some players might implement other equilibrium strategies \cite{SunGame2020}, which is reflected in the second challenge discussed above: uncertain behavior modes.
Consequently, the ego vehicle needs to hold a belief about the strategies employed by other vehicles and update the belief based on the observed trajectories. 
Common approaches \cite{tianAnytimeGameTheoreticPlanning2021, SunGame2020} in this domain often discretize the action space in line with the so-called Partially Observable Markov Decision Process (POMDP) \cite{lauriPartiallyObservableMarkov2023}.
To consider the continuous action space, branch model predictive control (BMPC) methods \cite{ChenBranchMPC2022, HardyContingencyPlanning2013} have been proposed, which utilize a trajectory tree with multiple branches to capture various potential behavior modes.
Compared to conventional robust motion planning \cite{AhnSafeMotionPlanning2022, WangNonGaussian2020}, only the shared nodes in the tree need to meet safety requirements across all possible scenarios, resulting in less conservative control inputs.

\subsection{Contribution}
To overcome the two main challenges, we propose a two-layer planning framework.
At the behavior planning layer, we approach the lane-merging problem in dense traffic from a game-theoretic perspective for interaction modeling.
The behavior planner computes multiple equilibrium strategies and outputs the corresponding multi-vehicle trajectories.
Next, we develop a motion planner in the framework of BMPC to refine the trajectories generated by the behavior planner and handle the various possible equilibrium strategies of other vehicles.
Our contributions with respect to the related literature are summarized as follows:
\begin{enumerate*}[label=(\roman*)]
\item 
We model the interaction between vehicles as a two-player matrix game, where one player is the ego vehicle and the second player represents all the surrounding vehicles. 
Unlike the optimization-based methods in \cite{cleachALGAMESFastSolver2020, FridovichILQG2020}, 
we introduce a matrix game as an approximation of the original dynamic game to circumvent the issue of local solutions.
To make the algorithm efficient, motivated by \cite{dingEPSILONEfficientPlanning2022, cunninghamMPDMMultipolicyDecisionmaking2015}, we approximate the action space of the dynamic game by using semantic-level actions and forward simulating the multi-vehicle system.
Moreover, inspired by POMDPs, we account for the unknown cost functions by incorporating the belief about the semantic-level actions.
\item 
We systematically evaluate the multi-vehicle trajectories by solving a matrix game.
A similar work EPSILON \cite{dingEPSILONEfficientPlanning2022} forward simulates the vehicle system, but it selects the optimal trajectory using handcrafted criteria.
Unfortunately, whenever the surrounding vehicles have potentially diverse behavior modes, the rule-based trajectory evaluation in EPSILON might result in overly aggressive or overly conservative trajectories \cite{LiMultipolicy2023}.

\end{enumerate*}

A preliminary version of this work was presented in \cite{LuyaoMaxtrixGame2023}. In this extended paper, we present the following additional contributions: 
\begin{enumerate*}[label=(\roman*), resume]
\item 
We develop a BMPC framework to address the strategy misalignment between the ego vehicle and other vehicles. On the contrary, some existing methods \cite{ZhangGameMPC2020, weiGameTheoreticMerging2022} simply focus on a single type of equilibrium without considering uncertainties in equilibrium strategies.

\item 
To achieve real-time performance, we implement a customized numerical solver that leverages the parallel structure of the trajectory tree and initialize the trajectory tree using our game-theoretic behavior planner.
Instead, most BMPC approaches presented in the literature \cite{ChenBranchMPC2022, WangInteraction2023, huActiveUncertaintyReduction2023a} rely on off-the-shelf nonlinear optimization solvers without exploiting the problem structure.

\item To validate the effectiveness of the proposed game-theoretic planning framework, we conduct extensive numerical simulations on real-traffic data, specifically, on the INTERACTION dataset \cite{interactiondataset2019}. 
Moreover, we compare our newly proposed planner with our previous one and other baseline methods.
\end{enumerate*}

\subsection{Related Work}
\subsubsection{Game-Theoretic Planning}
Game-theoretic methods have been applied in both motion planning and behavior planning.
The game-theoretic motion planners \cite{FridovichILQG2020, cleachALGAMESFastSolver2020} 
break the traditional predict-and-plan pipeline by jointly planning trajectories for all vehicles.
However, these methods can only find a local equilibrium, and the quality of the solution might heavily depend on the initial guess trajectories due to the nonlinear system dynamics and nonconvex constraints.
Instead, other approaches focus on the behavior planning problem, where strategies are represented by discrete semantic-level decisions.
The authors in \cite{liuInteractionAwareTrajectoryPrediction2023} design the strategies of the ego vehicle as motion primitives and use a leader-follower game to model the surrounding vehicles. 
Zhang et al. \cite{ZhangGameMPC2020} and Wei et al. \cite{weiGameTheoreticMerging2022} represent the strategy of the ego vehicle as the waiting time before merging and seek a Stackelberg equilibrium. 
In Stackelberg games, however, determining the relative role of the leader or follower might be difficult in practice \cite{liuInteractionAwareTrajectoryPrediction2023}. 
While most methods assume the ego vehicle to be the leader, the underlying rationale behind this choice is not always clear.
In contrast to the leader-follower structure, a Nash game treats all agents equally \cite{cleachALGAMESFastSolver2020}.
While there are methods based on Nash games \cite{lopezGameTheoreticLaneChangingDecision2022, HangHumanLike2021, CoskunRHMarkovGame2019}, our approach differs by incorporating interaction-aware forward simulation.
In addition to Nash and Stackelberg games, other approaches utilize level-\textit{k} reasoning to model human driving behavior \cite{tianAnytimeGameTheoreticPlanning2021, LiLevelKGame2022}. 
However, the computational burden of this framework is substantial due to the necessity of modeling the depth of human thinking \cite{jiHierarchicalGametheoreticDecisionmaking2023}.
Our game-theoretic planner uses semantic-level actions, falling into the category of behavior planning.

\subsubsection{POMDP}
The so-called Partially Observable Markov Decision Process (POMDP) provides a mathematical framework for handling incomplete information. 
One line of research is dedicated to approximating
original POMDP problems and enhancing computational
efficiency.
Online solvers, such as POMCP \cite{silverMonteCarloPlanningLarge2010}, POMCPOW \cite{sunbergImprovingAutomatedDriving2022a}, and DESOPT \cite{caiHyPDESPOTHybridParallel2021}, have been developed, which employ sampling techniques to estimate action-value functions. 
Other heuristic approximations of POMDPs have been introduced for specific autonomous driving applications to facilitate the search process. 
To generate interactive policies for lane-merging scenarios,
Hubmann et al. \cite{hubmannBeliefStatePlanner2018} combine the Monte Carlo sampling algorithm with an $\text{A}^*$ roll-out heuristic for fast convergence. 
Fischer et al. \cite{fischerGuidingBeliefSpace2022} propose an interactive lane-merging planner that utilizes trained policies to guide online belief planning. 
In \cite{HubmannAutomatedDriving2018, ZhouJointMultiPolicy2018}, the authors investigate the unsignalized intersection scenario and elaborately design the state and action space to achieve a low-dimensional problem that is solvable in practice.
The reader can refer to \cite{lauriPartiallyObservableMarkov2023} for a comprehensive survey on POMDP and its applications in robotics.
In our setting, we formulate a stochastic optimal control problem that is a continuous-time counterpart to POMDPs.

\subsubsection{Branch MPC}
BMPC is a specific instance of stochastic optimal control.
In autonomous driving applications, BMPC is used to account for the uncertain behavior modes of the surrounding vehicles.
Methods for BMPC vary in their approaches to modeling the surrounding vehicles.
The authors in \cite{qiuLatentBeliefSpace2020, huActiveUncertaintyReduction2023a} present a complete BMPC framework with belief update, enabling active information gathering. 
Chen et al. \cite{ChenBranchMPC2022} and Wang et al. \cite{WangInteraction2023} construct the trajectory trees with multiple branches; however, the probability attached to each branch depends only on the current joint state, rather than the observed historical trajectories.
By using interactive motion models, their methods still allow the ego vehicle to affect the motion of the surrounding ones. 
A further simplified version of BMPC is proposed in \cite{HardyContingencyPlanning2013, DaComprehensive2022, BatkovicRobustSMPC2021}, where the motion of the surrounding vehicles is determined by the multi-modal trajectories from a prediction module. 
These methods assume that both the future trajectory of each surrounding vehicle and the probability associated with each tree branch are fixed.
Our motion planner builds upon these simplified BMPC approaches. 
Since the behavior planner already accounts for mutual interactions by jointly planning trajectories for all vehicles, the motion planner, as a downstream module, disregards these interactions to achieve real-time performance.

The paper is structured as follows.
We introduce the lane-merging scenario and define the dynamic game problem in Section \ref{sec:problem_setting}. Section \ref{sec:GTBP} presents the game-theoretic behavior planner. 
In Section \ref{sec:trajectory_tree}, we first present a general formulation of the BMPC problem and subsequently elaborate on the design details. The numerical simulation results are showcased in Section \ref{sec:numerical_simulations}. 
Finally, Section \ref{sec:conclusion} concludes the paper.

\begin{figure}[t]
    \centering
    \includegraphics[width=\columnwidth]{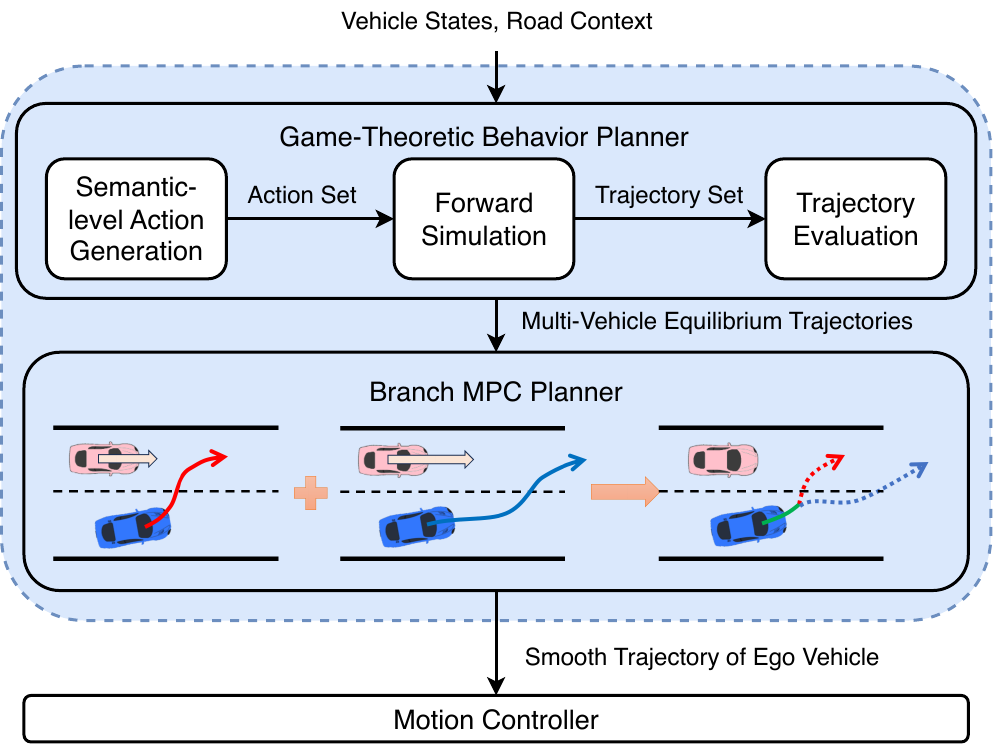}
    \caption{Structure of the proposed game-theoretic planner. 
    The behavior planner outputs the multi-vehicle trajectories that represent different equilibrium strategies.
    The BMPC planner utilizes the so-called trajectory tree to account for the potential behavior modes of the surrounding vehicles. 
    In this example, the separate tree branches handle two distinct behavior modes (equilibrium strategies) of the interacting vehicle (pink), whereas the shared part (green) needs to accommodate both potential behavior modes.}
    \label{fig:planner_structure}
    \vspace{-0.35cm}
\end{figure}

\section{Problem Setting} \label{sec:problem_setting}
\subsection{Lane-merging Scenario and General Planning Structure}
\begin{figure}\captionsetup[subfigure]{font=scriptsize}
    \centering
    \subfloat[]{%
        \includegraphics[width=0.4\textwidth]{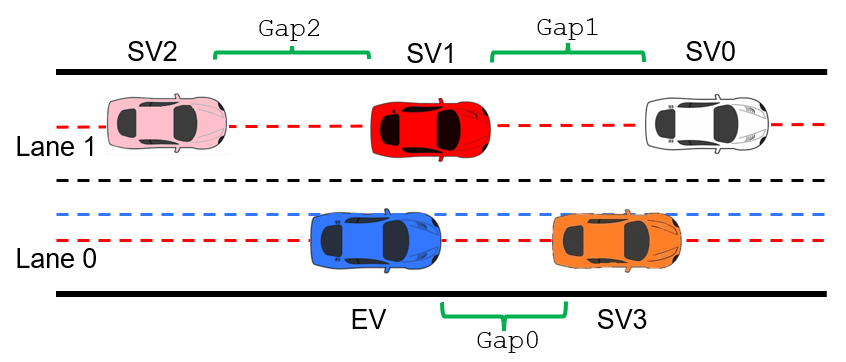}
        \label{fig:semantic_actions}
    }\\
    \subfloat[Interaction graph when EV selects \texttt{Gap1}.]{%
        \includegraphics[width=0.35\columnwidth]{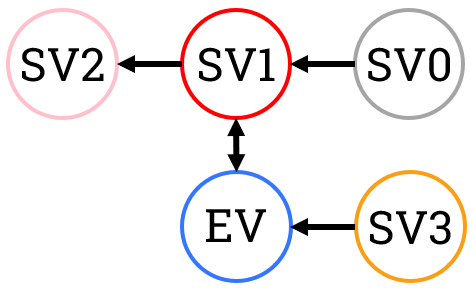}
        \label{fig:graph_gap1}
    }
    \hspace{3mm}
    \subfloat[Interaction graph when EV selects \texttt{Gap2}.]{%
        \includegraphics[width=0.35\columnwidth]{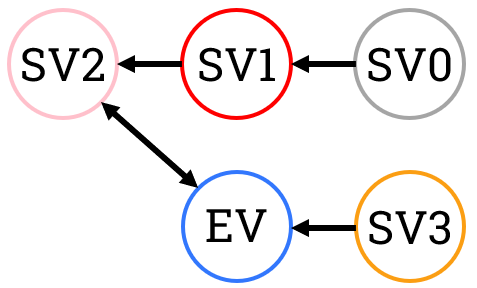}
        \label{fig:graph_gap2}
    }
    \caption{(a): The lane-merging problem: 
    three gaps are available for the ego vehicle to choose from: \texttt{Gap0}, \texttt{Gap1} and \texttt{Gap2}.
    The dashed red lines represent the centerlines of the lanes, and the dashed blue line represents the probing line.
    EV stands for the ego vehicle (blue), 
    while SV0, SV1, SV2 and SV3 denote the surrounding vehicles. (b)-(c): Illustrate the interaction graphes when EV selects \texttt{Gap1} and \texttt{Gap2}, respectively. The single-headed arrow indicates one-way influence, while double-headed arrows denote mutual interaction.}
    \vspace{-3mm}
\end{figure}

We consider a mixed-traffic scenario where an automated vehicle (ego vehicle) interacts with the surrounding vehicles, as shown in Figure \ref{fig:semantic_actions}.
The ego vehicle attempts to execute a lane change in response to the high-level command from either the human driver \cite{dingEPSILONEfficientPlanning2022} or the lane-selection planner \cite{bae2021risk}.
To this end, its decision-making system must account for various factors, including the merging gap, the timing for lane changing, and the diverse behaviors of the surrounding vehicles.
For example, in Figure \ref{fig:semantic_actions}, the ego vehicle can select to merge ahead of or after the SV1. 
If the SV1 yields, then the gap enlarges and the ego vehicle merges ahead of the SV1; otherwise, if the gap is not sufficiently wide, the ego vehicle might slow down and merge after the SV1. 

To tackle the lane-merging problem, we propose a game-theoretic planning framework, as illustrated in Figure \ref{fig:planner_structure}. 
Unlike traditional behavior planners that usually require a motion predictor as an upstream module, our approach combines motion prediction and behavior planning.
The proposed game-theoretic behavior planner comprises three core modules. 
Specifically, we begin by enumerating the possible semantic-level decision sequences.
For the ego vehicle, a semantic-level decision typically involves making a lane change, accelerating or decelerating. 
Then, we create the action pairs by combining the decision sequences of the ego vehicle and surrounding vehicles.
For each action pair, the forward simulator generates the multi-vehicle trajectories, and the trajectory evaluator computes the costs associated with these trajectories.
Finally, we construct a matrix game and seek multiple equilibrium strategies.
The trajectory evaluator outputs the corresponding multi-vehicle equilibrium trajectories.
Moreover, considering the uncertainty in the equilibrium strategies of surrounding vehicles,
we develop a BMPC planner to handle the multi-modal behaviors exhibited by these vehicles, as shown in Figure \ref{fig:planner_structure}. 
Specifically, since the ego vehicle (in blue) is uncertain whether the nearby vehicle (in pink) will yield, it plans a trajectory tree with a shared branch that accommodates both possible decisions of the pink vehicle. Beyond the branching point, we assume the ego vehicle has full knowledge of the pink vehicle's decision.


\subsection{Dynamic Game-Theoretic Setting} \label{subsec:matrix_game}
We model the vehicle interaction in the lane-merging problem using a finite horizon dynamic game with $N$ vehicles.
The dynamics of vehicle $i$ is represented by a kinematic bicycle model \cite[Section 2.2]{rajamaniVehicleDynamicsControl2011}:
\begin{align}
    \dot{p}_{x}^i = v^i\cos{(\theta^i)},\; \dot{p}_{y}^i = v^i\sin{(\theta^i)},\; 
    \dot{\theta}^i = \frac{v^i} {l^i}\tan(\delta^i),\; \dot{v}^i = a^i, \label{eq:bicycle}
\end{align}
where $(p_{x}^i, p_{y}^i)$, $\theta^i$ and $v^i$ are the position, heading angle, and speed, respectively; $a^i$ and $\delta^i$ are the acceleration and steering angle; $l^i$ represents the inter-axle distance. 
The state and control input vectors are denoted by $x^i := (p_{x}^i, p_{y}^i, \theta^i, v^i)$ and $u^i := (a^i, \delta^i)$, respectively.
Applying the Runge-Kutta method \cite[Section 8.2]{rawlings2017model},
we derive the following discrete-time dynamics:
\begin{equation} \label{eq:discrete dynamics}
    x_{t+1}^i  = f(x_{t}^i, u_{t}^i),
\end{equation}
where $t \in [0, T-1] := \{0,\dots,T-1\}$, $x_{t}^i \in \mathbb{R}^n$ and $u_{t}^i \in \mathbb{R}^m$ denote the state and control input vectors at time step $t$. 
For notational simplicity, the notation without a superscript refers to all vehicles, while the notation without $t$ refers to all time steps. 
For instance, $\boldsymbol{x}_t$ represents the states of all vehicles at time step $t$, while $x^i$ denotes the state trajectory of vehicle $i$. 
Moreover, $\boldsymbol{x}^{-i}$ and $\boldsymbol{u}^{-i}$ are the stacked state and control input vectors of all vehicles except vehicle $i$, respectively.
In this open-loop dynamic game, each vehicle selects a control input sequence to optimize its individual objective function. 
Thus, we formulate the $N$ interdependent optimization problems as follows:
\begin{subequations} \label{eq:game_formulation}
    \begin{empheq}[left={\empheqlbrace\,}]{align}
        \min_{u^i} \quad & J^i(\boldsymbol{x}, u^i, \boldsymbol{u}^{-i}) \label{eq:cost_game}\\
        \textrm{s.t.} \quad & x^i_{t+1} = f(x^i_t, u^i_t),\quad \forall t \in [0, T-1], \label{eq:dynamics_game} \\
        & u^i_t \in \mathcal{U}, \quad \forall t \in [0, T-1],
  \end{empheq}
\end{subequations}
where $\mathcal{U} \subseteq \mathbb{R}^m$ denotes the feasible control input set, representing the bounds for acceleration and steering angle, $T$ is the horizon and the cost function $J^i$ typically measures several user-defined metrics, such as progress, comfort, and safety. It depends on both the joint state trajectory and the control input sequences for each vehicle.
The solution to \eqref{eq:game_formulation} is a Nash equilibrium, where no vehicle can further reduce its cost by unilaterally changing its control input sequence.
Unfortunately, directly solving \eqref{eq:game_formulation} is challenging due to the nonconvex cost function \eqref{eq:cost_game} and the nonlinear dynamics \eqref{eq:dynamics_game}. 
Consequently, if the initial guess is relatively far from an equilibrium point, the solution is likely to get stuck at an undesirable point.
To mitigate the aforementioned issues, we use the search-based method, which can explore the solution space more extensively. 
Specifically, we first discretize the control input space in \eqref{eq:game_formulation} and then formulate a matrix game with discrete action space.  
A matrix game can be defined by a tuple $(\mathcal{N},\Pi,J)$, where $\mathcal{N}$ is the set of players, $\Pi = \times_{i\in \mathcal{N}} \Pi^i$ is the joint action space, and 
$J = \left(J^i \right)_{i\in \mathcal{N}}$ 
are the cost functions. 
In a matrix game, we seek equilibria of the following types.
\begin{definition}
(Pure-strategy Nash equilibrium). 
A pure-strategy Nash equilibrium is a set of players' actions, $\{\pi^{i*}\}_{i\in\mathcal{N}}$ such that, for each player $i$, it holds that
$$
J^i(\pi^{i*},\,\pi^{-i*}) \leq \inf_{s^i \in \Pi^i } J^i(s^i,\,\pi^{-i*}),
$$
where $\pi^{-i}$ represents the set of actions taken by all players except player $i$.
\end{definition}
\smallskip

\begin{definition}
(Stackelberg equilibrium).  
A Stackelberg equilibrium is a pair 
$\{\pi^{\textrm{L}*},\,\pi^{\textrm{F}*}(\cdot)\}$ 
such that
\begin{align*}
    \pi^{\textrm{L}*} = \mathrm{arg}\min_{\pi^{\textrm{L}} \in \Pi^{\textrm{L}}} J^{\textrm{L}}(\pi^{\textrm{L}},\, \pi^{\textrm{F}*}(\pi^{\textrm{L}})), \\
    \pi^{\textrm{F}*}(\pi^{\textrm{L}}) = \mathrm{arg}\min_{\pi^{\textrm{F}} \in \Pi^{\textrm{F}}} J^{\textrm{F}} (\pi^{\textrm{L}},\,\pi^{\textrm{F}}),
\end{align*}
where the superscripts, \textrm{L} and \textrm{F}, represent the leader and the follower of the game, respectively.
\end{definition}
\noindent The idea behind the Stackelberg game is that the leader takes action first, and then the follower plays the best response \cite{baar_dynamic_1998}. 
The roles of leader and follower are not always fixed on the road. In other words, the EV can switch between being the leader and the follower \cite{wang_social_2022}. 
Thus, we can consider two Stackelberg equilibria: one with the EV as the leader and the other with the EV as the follower.

The proposed matrix game can be viewed as a 
discrete approximation of \eqref{eq:game_formulation}. Compared with solving the origin problem, the solution to the matrix game is easy to compute, e.g. by enumeration.
Moreover, owing to the context-aware action generation and interactive forward simulation (see Section \ref{sec:GTBP}), the problem size is moderate. 

\section{Game-Theoretic Behavior Planning} \label{sec:GTBP}
In this section, we provide the technical details of the matrix game, covering aspects such as players, action generation, forward simulation, and trajectory evaluation. 

\subsection{Game Players}
In principle, we need to consider both the ego vehicle and all relevant surrounding vehicles as players. This choice accounts for the interaction between each pair of vehicles but results in a complex multi-player matrix game.
To simplify the matrix game, we make several assumptions based on the inherent properties of the lane-merging problem.
\begin{assumption} \label{assumption:one_interacting_vehicle}
For each semantic-level decision, the ego vehicle mutually interacts with at most one surrounding vehicle.
All other vehicle interactions are considered unidirectional.
\end{assumption}
This assumption is illustrated in Figure \ref{fig:graph_gap1} and \ref{fig:graph_gap2}. 
Specifically, in Figure \ref{fig:graph_gap1}, the EV does not influence the vehicle (SV0) ahead of the SV1, and the SV2 may be affected by the SV1 but does not influence the EV.
It is important to note that Assumption \ref{assumption:one_interacting_vehicle} does not imply that our approach considers only one surrounding vehicle as the interacting vehicle (IV). 
Instead, conditioned on the semantic-level decision, we select the appropriate IV, as shown in Figure \ref{fig:graph_gap1} and \ref{fig:graph_gap2}. 
Two potential interacting vehicles are considered in this example.
Solving the game allows the EV to decide on the merging gap and the corresponding IV.
Moreover, inspired by RSS \cite{shalev-shwartz_formal_2018}, we assume that the SVs with one-way interactions always take action to avoid colliding with the lead vehicle. Combining this with Assumption \ref{assumption:one_interacting_vehicle}, we can treat the SVs as a vehicle group (VG). 
In the group, the IV owns multiple semantic-level actions, while the remaining SVs exhibit car-following behavior.
To sum up, we consider the EV and the group of SVs (VG) as two players, $\mathcal{N} := \{ \rm{EV}, \rm{VG} \}$.

\begin{assumption} \label{assumption:maintain_lanes}
In the behavior planning problem, the surrounding vehicles maintain their lanes and move longitudinally
\cite[Section 2]{liuInteractionAwareTrajectoryPrediction2023}, \cite[Section 3]{weiGameTheoreticMerging2022}.
\end{assumption}

\subsection{Semantic-Level Action Generation} \label{subsec:scenario_generation}
One straightforward choice for the action is the control input sequence of each player, which can be obtained by discretizing the control input space.
However, this approach results in large action sets and a large-scale matrix game that is potentially challenging to solve. 
Moreover, if we independently sample the control input space for each player, the potential interaction between players would be ignored, and the resulting trajectories might not be realistic. 
Inspired by human drivers, we instead represent the action of player $i \in \mathcal{N}$ by a semantic-level decision sequence, denoted as $\pi^i := \{\pi_{0}^i,\dots,\pi_{k}^i,\dots,\pi_{H-1}^i\}$, where $H$ is the decision horizon.
Following this, we can now discuss the actions of the ego vehicle and surrounding vehicles.

\subsubsection{Actions of Ego Vehicle}
In the lane-merging problem, the semantic-level decision involves selecting a gap and determining a desired lateral position. 
For example, in Figure \ref{fig:semantic_actions}, the EV has three potential gaps to choose from. 
To reach the target gap, the EV needs to perform a sequence of lateral decisions. The common lateral decisions are lane changing and lane keeping. 
Additionally, we introduce one additional intermediate lane, represented by the dashed blue line in Figure \ref{fig:semantic_actions}, to enable a probing decision. 
This allows the EV to gather information and negotiate with the SVs.
Overall, the complete lateral decision set can be defined as:
\begin{align*}
    D^{\textrm{lat}} := \{\texttt{LaneKeep},
    \texttt{LeftChange},
    \texttt{LeftProbe}\}.
\end{align*}
The semantic-level decision at decision step $k$ is denoted by an action pair $ \pi^{\textrm{EV}}_k := (g_k, d^{\textrm{lat}}_k)$, where $g_k \in \{\texttt{Gap0},\, \texttt{Gap1},\, \texttt{Gap2}\}$ and  $d^{\textrm{lat}}_k \in D^{\textrm{lat}}(g_k)$.
We note that the lateral decision set is conditioned on the gap selection, which reduces the number of action pairs. 
For example, if the EV chooses \texttt{Gap0}, then the only available lateral action is to keep the current lane.

Then, in line with \cite{dingEPSILONEfficientPlanning2022}, we construct a decision tree to enumerate all possible decision sequences. 
Each node in the tree represents a decision pair. The decision tree is rooted in the decision selected in the last planning cycle and branches out at each decision step. 
Due to the exponential growth of the number of decision sequences with the depth of the tree, it is necessary to prune the decision tree to limit computational complexity. By using semantic-level decisions, we can design some rules to prune the tree. For instance, we can restrict the number of decision changes over the planning horizon because human drivers tend to maintain their current driving decisions for relatively long periods. In addition, we can rule out certain transitions that would not be considered by normal human drivers, such as the transition from $(\texttt{Gap1}, \texttt{LeftChange})$ to $(\texttt{Gap2}, \texttt{LeftChange})$.

\subsubsection{Actions of the Surrounding Vehicles} \label{subsubsec:actions_surrounding_vehicles}
Next, let us postulate another working assumption.
\begin{assumption} \label{assumption:unchanged_action}
In the behavior planning problem, the surrounding vehicles keep their decisions unchanged throughout each forward simulation.
\end{assumption}
\noindent Even though this assumption limits the decision space, it remains reasonable in practice since the behavior planner runs in a receding horizon fashion.
In view of Assumptions \ref{assumption:maintain_lanes} and \ref{assumption:unchanged_action}, we define the action set of the VG as $\{\texttt{Assert}, \texttt{Yield}\}$. 
Assumption \ref{assumption:one_interacting_vehicle} indicates that the IV selects an action from the set $\Pi^{\textrm{VG}} := \{ \texttt{Assert}, \texttt{Yield} \}$ while the other vehicles in the group adhere to car-following behavior. 
For instance, in Figure \ref{fig:semantic_actions}, if the EV selects \texttt{Gap1}, then it mutually interacts with the SV1, and the SV2 just keeps a safe distance from the SV1.


\begin{figure}[t!]
    \vspace{0mm}
    \centering
    \includegraphics[width=0.43\textwidth]{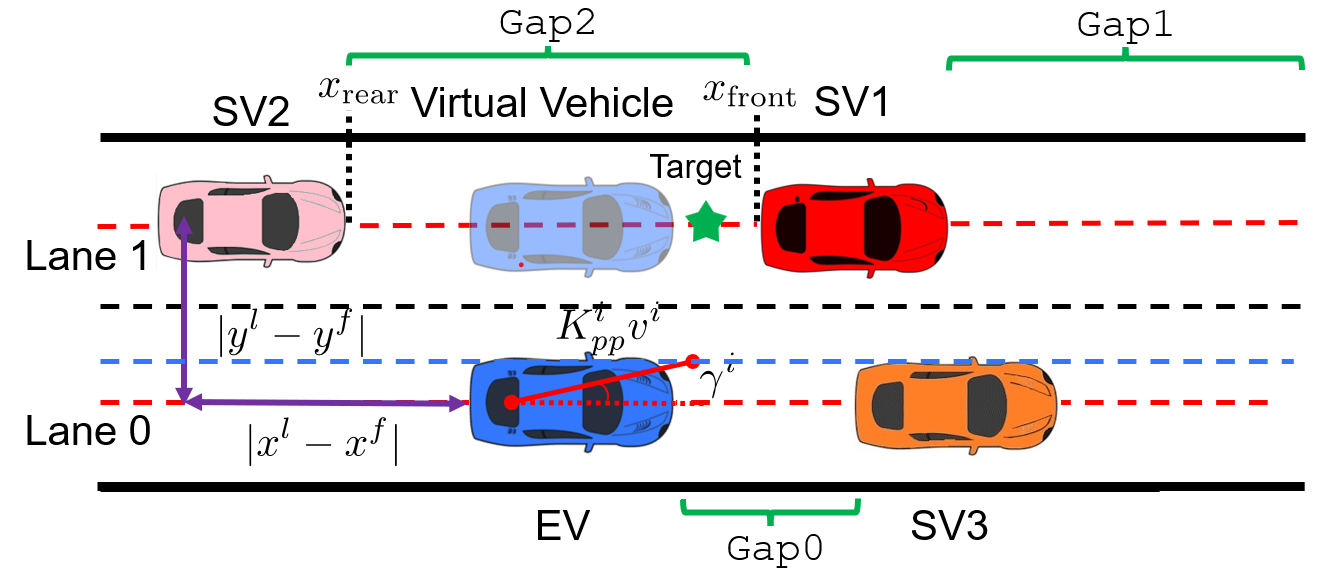}
    \caption{Multi-vehicle forward simulation.   
    $|x^l-x^f|$ and $|y^l-y^f|$ represent the longitudinal and lateral distances between the lane-changing vehicle (leader) and the interacting vehicle (follower), respectively.}
    \label{fig:controllers}
    \vspace{-3mm}
\end{figure}

\subsection{Multi-vehicle Forward Simulation} 
We generate trajectories for multiple vehicles by simulating their motion from the initial states. Their states evolve according to \eqref{eq:discrete dynamics}. 
Next, we discuss the computation of control inputs considering vehicle interactions.
\label{subsec:forward_simulation}

\subsubsection{Motion of the Ego Vehicle}
We apply two separate controllers to generate the longitudinal and lateral motion for the EV.
For longitudinal motion, we track the target longitudinal position and the desired speed using a PD controller \cite[Section 7.B]{dingEPSILONEfficientPlanning2022}.
\textcolor{black}{One example of} the target longitudinal position within the desired gap is illustrated in
Figure \ref{fig:controllers}. 
The desired gap is defined based on the positions of the front and rear vehicles, denoted as $x_{\text{front}}$ and $x_{\text{rear}}$, respectively.
Similarly to \cite[Section 7.B]{dingEPSILONEfficientPlanning2022}\cite[Section 3]{shalev-shwartz_formal_2018}, we determine the target longitudinal position and speed based on the safe distance that the ego vehicle and the rear vehicle (shown in pink in Figure \ref{fig:controllers}) need to keep.
The output of the PD controller is represented by $a_{\text{pd}}$. However, this controller does not consider the lead vehicle and the static obstacle. 
Hence, to avoid a collision, we compute the car-following acceleration $a_{\text{idm}}$ using the intelligent driver model (IDM) \cite{treiber_congested_2000}. 
Overall, for the sake of safety, we select the minimum acceleration as the command, $a_{\text{cmd}} = \min(a_{\text{pd}}, a_{\text{idm}})$.

As for lateral motion, we adopt a pure pursuit controller 
that takes the current vehicle speed and the target line as inputs. The steering angle is computed by
$
    \delta_{\textrm{cmd}}^i = \tan^{-1}\left(2l^i\sin(\gamma^i) / (K^i_{pp}v^i)\right),
$
where $\gamma^i$ represents the angle between the heading direction and lookahead direction, $K_{pp}$ is the feedback gain, and $K^i_{pp}v^i$ is the lookahead distance. 

\subsubsection{Motion of the Surrounding Vehicles}
\label{sec:non_ego_in_planner}
To model car-following behavior and reactions to lane-changing vehicles,  we propose a modified IDM.
This model considers lane-changing vehicles by projecting them onto their target lanes, resulting in virtual vehicles as shown in Figure \ref{fig:controllers}.
The only difference between the standard IDM and our modified version lies in the computation of the distance between the leader and the follower, which is given by:
\begin{align}
    d_{\textrm{idm}} = |x^l - x^f| e^{\kappa |y^l - y^f|}, \; \kappa = 2\log(\beta)/w_{\textrm{lane}}, \label{eq:idm_distance}
\end{align}
where $w_{\textrm{lane}}$ is the lane width, and $\beta$ is a parameter characterizing the level of willingness to yield.
By adjusting the value of $\beta$, we can represent different actions performed by the VG. 
Specifically, a large value of $\beta$ indicates that the vehicle on the target lane is less likely to yield to the lane-changing vehicle because it perceives that the projection is far away. 
When the lateral distance between two vehicles vanishes, that is $|y^l-y^f|=0$, the virtual distance between them is equivalent to the actual longitudinal distance. 

\subsection{Trajectory Evaluation} \label{subsec:trajectory_evaluation}
\subsubsection{Cost Function}
Once multi-vehicle trajectories are generated, we proceed to select a specific action pair by solving a matrix game.
To construct the cost matrix, we introduce the cost function $J^i$ of vehicle $i$, which is a combination of several user-defined metrics, including safety, efficiency, comfort, navigation, and information cost: $J^i = J_{\textrm{saf}}^i + J_{\textrm{eff}}^i + J_{\textrm{com}}^i + J_{\textrm{nav}}^i + J_{\textrm{inf}}^i$.
The value of the cost function $J^i$ depends on the simulated trajectories, which are influenced by the semantic-level action of the EV, $\pi^{\textrm{EV}}$, and the VG, $\pi^{\textrm{VG}}$.
We consider the SVs as a vehicle group by calculating the total cost as 
$J^{\textrm{VG}} := \sum_{i \neq \text{EV}} J^i$.

We calculate the safety cost by examining vehicle collisions.
The footprint of vehicle $i$ is modeled as a rectangle.
If the distance $d^{ij}(t)$ between two rectangles is less than a small value $\underline{d}$, indicating a potential collision, we assign a substantial penalty $w_1^{\text{saf}}$ to the corresponding trajectory. 
Additionally, we encourage the vehicle to keep a suitable distance from its surroundings by giving a relatively small penalty $w_2^{\text{saf}}$ when $d^{ij}(t)$ falls within the range of $[\underline{d},\bar{d}]$. 
Next, we measure the efficiency of the trajectory by computing the sum of the squares of the differences between the vehicle speed and its desired speed: 
$
J_{\textrm{eff}}^i(\pi^{\textrm{EV}}, \pi^{\textrm{VG}}):= w^{\textrm{eff}}\sum_{t=0}^{T} (v^{i}(t) - v^i_\text{des})^2.
$
For the comfort cost, we consider the acceleration change, commonly called jerk.
We utilize the finite difference to approximate the jerk, and subsequently define the comfort cost:
$
J_{\textrm{com}}^i(\pi^{\textrm{EV}}, \pi^{\textrm{VG}}) := w^{\textrm{com}}\sum_{t=1}^{T} (a^{i}(t)-a^{i}(t-1))^2/\Delta t^2.
$
Next, we penalize the differences between the vehicle's lateral position and its desired lateral position to encourage the lane-changing maneuver:
$
    J_{\textrm{nav}}^i(\pi^{\textrm{EV}}, \pi^{\textrm{VG}}) := w^{\textrm{nav}}\sum_{t=0}^{T} (y^{i}(t) - y^i_{\textrm{des}})^2.
$
Additionally, we introduce an information gain metric in the cost function, inspired by \cite{sadighPlanningCarsThat2018}, to motivate the ego vehicle to actively identify the intentions of other vehicles.
Informally speaking, the information cost encourages the ego vehicle to take the \texttt{Probing} action.

\subsubsection{Belief Update}
The EV needs to estimate the cost functions of the SVs by observing their trajectories since direct access to these cost functions is impractical.
Inspired by POMDP, we account for uncertainty in the aggregate cost of the SVs by   
integrating the beliefs into the cost function.
The modified aggregate cost is computed as follows:
$$
    \bar{J}^{\textrm{VG}}_{ij} := \left(1-b\left(\pi^{\textrm{VG}}_i\right)\right) J^{\textrm{VG}}_{ij},\quad \sum_{i=1}^{M^{\textrm{VG}}} b\left(\pi^{\textrm{VG}}_i\right) = 1,
$$
where 
$i$ and $j$ represent the indices in the cost matrix, 
$\pi^{\textrm{VG}}_i \in \Pi^{\textrm{VG}}$ is the action of the VG, $M^{\textrm{VG}}:= \left\vert \Pi^{\textrm{VG}} \right\vert$ denotes the cardinality of the action set of the VG, and $b$ represents the belief about the VG's action.
This design can be understood as incorporating prior knowledge about the behavior of the VG into the aggregate cost. 
For example, if we have prior knowledge suggesting that the VG is inclined to yield, then we can set the corresponding belief close to $1$, which reduces the aggregate cost.

The belief is recursively updated at the beginning of each planning cycle using Bayes filtering \cite{thrunProbabilisticRobotics2002}. 
The update rule involves two key components: 
the transition model 
$\mathbb{P}\left(\pi^{\text{VG}} \mid \pi^{\text{VG}}_{-}\right)$ 
and the observation model 
$\mathbb{P}\left(o \mid \pi^{\text{VG}}\right)$, 
where 
the notation $\square_{-}$ is associated with the previous planning cycle, and 
$o$ denotes the observed state. 
Assuming that the interacting vehicle in the group does not change its action, the transition model can be significantly simplified.
In the simplified model, 
$\mathbb{P}\left(\pi^{\text{VG}} = \pi^{\text{VG}}_j \mid \pi^{\text{VG}}_i\right)$ is equal to $1$ if $i$ and $j$ coincide,
and for all other cases, 
$\mathbb{P}\left(\pi^{\text{VG}} = \pi^{\text{VG}}_j \mid \pi^{\text{VG}}_i\right)$ is equal to $0$.

In the observation model, we use the dynamics \eqref{eq:discrete dynamics} with additive Gaussian noise to predict the IV's state in the group:
\begin{align*}
    x^{\text{VG}}_{t+1} = f\left(x^{\text{VG}}_t, u^{\text{VG}}_t\left(\boldsymbol{x}_t, \pi^{\text{VG}}_t, \pi^{\text{EV}}_t\right)\right) + w,\; w \sim \mathcal{N}(0, W),
\end{align*}
where $u^{\text{VG}}_t(\cdot)$ denotes the control policy discussed in Section \ref{sec:non_ego_in_planner} and $w$ is additive Gaussian noise with zero mean and covariance matrix $W$. 
Then, we compute the likelihood of receiving 
an observation $o$ using the following distribution:
\begin{align*}
    \mathbb{P}\left(o \mid \pi^{\text{VG}}\right) \sim \mathcal{N}\left(f\left(x^{\text{VG}}_{-}, u^{\text{VG}}_{-}\left(\boldsymbol{x}_{-}, \pi^{\text{VG}}_{-}, \pi^{\text{EV}}_{-}\right)\right), W\right),
\end{align*}
where the mean is the predicted state. With the transition model and the observation model, the belief update rule can be expressed as follows:
\begin{align*}
    b(\pi^{\text{VG}}) = \frac{\mathbb{P}\left(o \mid \pi^{\text{VG}}\right) \sum\limits_{\pi^{\text{VG}}_{-}} \mathbb{P}\left(\pi^{\text{VG}} \mid \pi^{\text{VG}}_{-}\right) b\left(\pi^{\text{VG}}_{-}\right)}{\sum\limits_{\pi^{\text{VG}}} \mathbb{P}\left(o \mid \pi^{\text{VG}}\right)\sum\limits_{\pi^{\text{VG}}_{-}} \mathbb{P}\left(\pi^{\text{VG}} \mid \pi^{\text{VG}}_{-}\right) b(\pi^{\text{VG}}_{-})}.
\end{align*}

\subsubsection{Equilibrium Selection}
With the cost matrix, we compute multiple equilibria, such as Nash and Stackelberg equilibria, by enumerating all possible combinations of semantic-level actions.
If multiple Nash equilibria exist, we select the equilibrium with the lowest social cost \cite{zanardiUrbanDrivingGames2021}, defined as $\mathcal{C}_{ij} := J^{\text{EV}}_{ij} + J^{\text{VG}}_{ij}$.
If a pure-strategy Nash equilibrium does not exist, we choose the Stackelberg equilibrium \cite{ZhangGameMPC2020, weiGameTheoreticMerging2022} with the EV as the follower as a backup solution.



\section{Branch Model Predictive Control} \label{sec:trajectory_tree}
The motion planner aims at refining the coarse trajectory generated by the behavior planner.
We model the motion planning problem as a stochastic optimal control problem due to the unknown behavior modes of the surrounding vehicles. 
In this section, we first introduce the vehicle modeling and then the general formulation of the interaction-aware motion planning problem.
Subsequently, we introduce some simplifications to rapidly compute an approximate solution.

\subsection{Vehicle Modeling}
\subsubsection{Dynamics}
For simplicity, we consider two vehicles in this section: the ego vehicle (EV) and one interacting vehicle (IV), with the states denoted as $x^{\text{EV}}$ and $x^{\text{IV}}$, respectively. The joint state is represented as $\boldsymbol{x} := \left( x^{\text{EV}}, x^{\text{IV}} \right)$.
We model the vehicle dynamics by a discrete-time kinematic bicycle,
where $i \in \left\{ \text{EV}, \text{IV} \right\}$, $x^i := (p_{x}^i, p_{y}^i, \theta^i, v^i)$ and $u^i := (a^i, \delta^i) \in \mathcal{U}^i$ as in \eqref{eq:discrete dynamics}. $\mathcal{U}^i$ represents the feasible control input set.

\subsubsection{Behavior Model}
The interactive behavior of the IV can be modeled using the feedback policy $\kappa: \mathbb{R}^n \times \Omega \rightarrow \mathbb{R}^m$, where $\Omega$ is a finite set with each element corresponding to one behavior mode of the IV. The control input of the IV at time step $t$ is determined by both the joint state $\boldsymbol{x}_t$ and the behavior mode $\omega_t \in \Omega$, that is,
\begin{equation}
    \textcolor{black}{u_{t}^\text{IV} = \kappa(\boldsymbol{x}_t, \omega_t)}, \quad \omega_t \sim b_t := \mathbb{P}(\omega_t \mid O_t), \label{eq:FB_policy}
\end{equation}
where $b_t$, known as the belief state, is the distribution of $\omega_t$ conditioned on the observation $O_t$. The observation is a collection of the observed joint states $O_t := \left[\boldsymbol{x}_t, \boldsymbol{x}_{t-1},\dots, \boldsymbol{x}_0 \right]$. 
Similarly to \cite{liuInteractionAwareTrajectoryPrediction2023}, 
the EV updates the belief state based on Bayes filtering after obtaining a new observation. The belief update, also referred to as belief dynamics \cite{huActiveUncertaintyReduction2023a}, can be briefly expressed as:
$
    b_{t+1} = g\left(b_t, \boldsymbol{x}_{t+1}, u_{t}^\text{EV}\right).
$
In general, it is commonly believed that forward propagating the belief state analytically is intractable due to the nonlinear vehicle dynamics \cite{huActiveUncertaintyReduction2023a, qiuLatentBeliefSpace2020}. 
Therefore, some approximations become necessary in practice, as shown in Section \ref{subsec:trajectory_tree}.

\subsection{General Formulation}
With all these ingredients, we can now formulate the interaction-aware motion planning problem in the framework of stochastic optimal control:
\begin{subequations} \label{eq:general_formulation}
\begin{align} 
    \min_{\left(\mu_{t}^{\text{EV}}\right)_{[0, T-1]}} \quad & \textcolor{black}{
    \mathbb{E}_{\substack{\omega_t \sim b_t, \\ t \in [0, T-1]}}}
    \left\{ \sum_{t=0}^{T-1} \ell(\boldsymbol{x}_t, u_{t}^{\text{EV}}) + \ell_f(\boldsymbol{x}_T) \right\} \\
    \text{s.t.} \quad & \boldsymbol{x}_0 = \bar{\boldsymbol{x}},\; b_0 = \bar{b}, \\ 
    & \forall t \in [0, T-1]: \nonumber \\
    & x_{t+1}^{\text{EV}} = f(x_{t}^{\text{EV}}, u_{t}^{\text{EV}}),\; \textcolor{black}{u_{t}^\text{EV} = \mu^{\text{EV}}_{t}(\boldsymbol{x}_t, b_t)}, \\
    & x_{t+1}^{\text{IV}} = f(x_{t}^{\text{IV}}, u_t^{\text{IV}}),\; u_{t}^\text{IV} = \kappa(\boldsymbol{x}_t, \omega_t), \\
    & b_{t+1} = g(b_t, \boldsymbol{x}_{t+1}, u_{t}^{\text{EV}}), \\
    &u_{t}^{\text{EV}} \in \mathcal{U}^{\text{EV}}, \label{eq:feasible_control_set} \\
    &\forall t \in [0, T]:\; 
    h(\boldsymbol{x}_t) \leq 0 ,
    \label{eq:collision_avoidance}
\end{align} 
\end{subequations}
where $\left(\mu_t^{\text{EV}}\right)_{[0, T-1]}$ is a sequence of feedback control policies, 
$\bar{\boldsymbol{x}}$ and $\bar{b}$ are the initial joint state and belief, 
$\mathcal{U}^{\text{EV}} \subseteq \mathbb{R}^m$ is the set of feasible control inputs, and \eqref{eq:collision_avoidance} represents the collision avoidance constraints. 
We defer the discussion on the feasible control input set \eqref{eq:feasible_control_set} and safety constraint \eqref{eq:collision_avoidance} to Section \ref{subsec:details}.
An alternative to hard collision avoidance constraints is a chance constraint formulation, which ensures safety in a probabilistic manner and reduces conservatism at the price of higher computational effort \cite{WangInteraction2023, AhnSafeMotionPlanning2022}.
Thus, to achieve a balance between robustness and computational demands, we adopt soft collision avoidance constraints in practice \cite{liniger_optimization-based_2015} and in turn sacrifice strict probabilistic guarantees. 

\begin{figure}[!t]
\centering
\includegraphics[width=0.9\columnwidth]{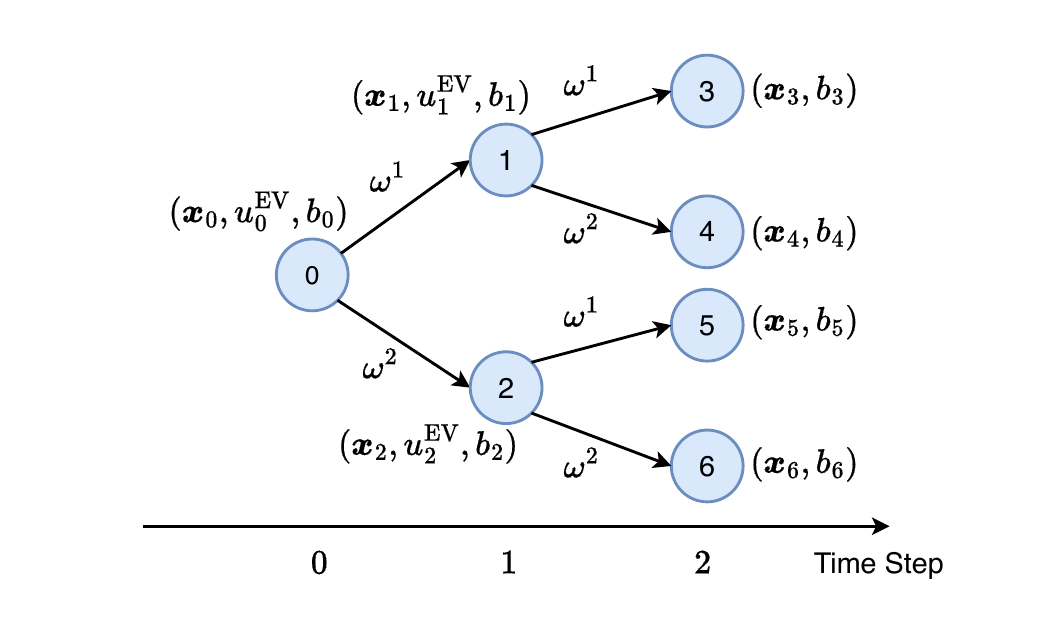}
\caption{Trajectory tree with a horizon of $T = 2$. 
The tree branches out at each time step based on the behavior mode $\omega_t \in \Omega:= \{\omega^1, \omega^2\}$. 
A transition probability $P_i$ pertains to each branch. 
In this example, $\mathcal{N} = \{0,1,2,3,4,5,6\}$ and $\mathcal{L} = \{3,4,5,6\}$.}
\label{fig:trajectory_tree}
\vspace{-0.3cm}
\end{figure}

\subsection{Trajectory Tree} \label{subsec:trajectory_tree}
Now, let us present an approximation of the general interaction-aware motion planning problem. 
In a numerical optimal control problem, we often look for a sequence of control inputs $\left( u_t^{\text{EV}} \right)_{[0, T-1]}$ rather than a sequence of feedback policies $\left(\mu_t^{\text{EV}}\right)_{[0, T-1]}$ to reduce the computational complexity. 
Nevertheless, a sequence of control inputs that solves \eqref{eq:general_formulation} might be overly conservative or even not exist in some cases due to the requirement that constraints must be respected across all realizations of uncertainty.
Another primary drawback of this approximation is neglecting the advantages that can be gained from future observations.
To resolve this, we instead seek a trajectory tree where all possible realizations of uncertainty are enumerated. 
As shown in Figure \ref{fig:trajectory_tree}, the tree is rooted at the node of the current joint state and branches out at each time step based on the behavior mode. 
We represent the set of all tree nodes as $\mathcal{N}$ and the set of all leaf nodes as $\mathcal{L}$. 
The transition probability from a parent node $\text{p}(i)$ to its child node is denoted by $P_i$.
In contrast to a single control sequence, distinct control inputs at each predicted time step are derived based on different behavior modes. 
Using the trajectory tree, we reformulate the stochastic optimal control problem \eqref{eq:general_formulation} as follows:
\begin{subequations} \label{eq:simplified_formulation}
\begin{align} 
    \min_{\left( u_i^{\text{EV}} \right)_{i \in \mathcal{N} \backslash \mathcal{L}}} &\sum_{i \in \mathcal{L}} P_i \ell_f\left(\boldsymbol{x}_i\right) + 
    \sum_{i \in \mathcal{N} \backslash \mathcal{L}} P_i \ell\left(\boldsymbol{x}_i, u_i^{\text{EV}}\right) \label{eq:tree_cost}\\
    \text{s.t.}\quad & \boldsymbol{x}_0 = \boldsymbol{\bar{x}},\; b_0 = \bar{b}, \\
    & \forall i \in \mathcal{N} \backslash \{0\}: \\
    & x_{i}^{\text{EV}} = f(x_{\text{p}(i)}^{\text{EV}}, u_{\text{p}(i)}^{\text{EV}}), \label{eq:tree_ev_dyn}\\
    & x_{i}^{\text{IV}} = f(x_{\text{p}(i)}^{{\text{IV}}}, u_{\text{p}(i)}^{\text{IV}}),\; u_{\text{p}(i)}^{\text{IV}} = \kappa(\boldsymbol{x}_{\text{p}(i)}, \omega_{\text{p}(i)}), \label{eq:tree_iv_dyn} \\
    & b_i = g(b_{\text{p}(i)}, \boldsymbol{x}_i, u_{\text{p}(i)}^{\text{EV}}), \label{eq:belief_simplified_formulation}\\
    & \forall i \in \mathcal{N} \backslash \mathcal{L}:\; u_{i}^{\text{EV}} \in \mathcal{U}^{\text{EV}}, \label{eq:tree_control_constr} \\
    & \forall i \in \mathcal{N}:\; h(\boldsymbol{x}_i) \leq 0, \label{eq:tree_state_constr}
\end{align}
\end{subequations} 
where objective function \labelcref{eq:tree_cost} computes the expected cost, and constraints \labelcref{eq:tree_ev_dyn,eq:tree_iv_dyn,eq:belief_simplified_formulation} describe the evolution of the states and belief state between two nodes. Constraints \labelcref{eq:tree_control_constr,eq:tree_state_constr} are associate with control limits and safety constraints.

\begin{figure}
    \centering
    \includegraphics[width=0.9\columnwidth]{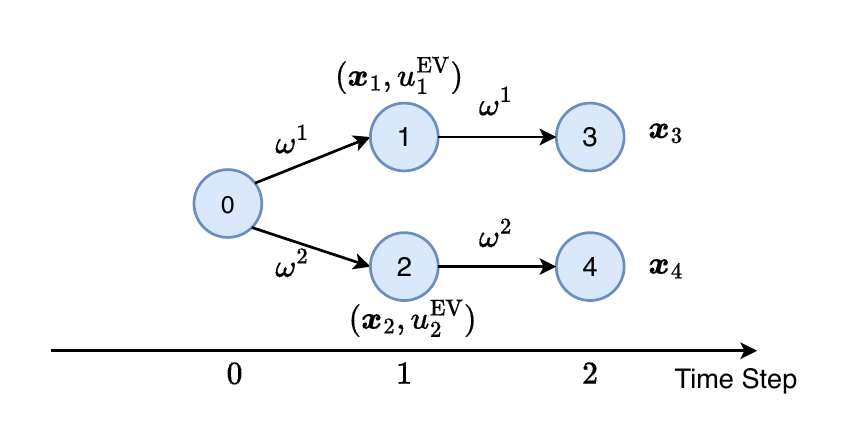}
    \caption{Simplified trajectory tree. The tree has multiple branches exclusively at the root node, while all other nodes, except for the leaf nodes, have only one single branch.}
    \label{fig:simple_trajectory_tree}
\end{figure}

While the trajectory tree in \eqref{eq:simplified_formulation} is a reasonable approximation of a sequence of feedback policies, it still entails several practical challenges. 
First, the introduction of additional optimization variables results in an increased computational burden. Additionally, as the optimization problem is nonconvex, the quality of the computed solution is heavily dependent on the initial guess.
Providing an initial trajectory tree is also a nontrivial task. 
Taking the aforementioned facts into account, we further simplify the trajectory tree as follows: 
\begin{enumerate*}[label=(\roman*)]
    \item The trajectory tree only branches out once, which assumes that the behavior mode of the IV remains constant over the planning horizon and the EV is fully certain about the behavior mode of the IV after the first branch. 
    As a result, we do not consider the belief update in \eqref{eq:belief_simplified_formulation} over the planning horizon. 
    Additionally, our behavior planner can easily initialize such a simplified tree, enhancing the quality of the computed solution. 
    \item The behavior planner provides the multi-modal state trajectories of the IV, which can be viewed as an open-loop control policy. 
    Compared with using feedback control policy in \eqref{eq:FB_policy}, we ignore the mutual interaction in motion planning since it has been taken into account by the behavior planner.
\end{enumerate*}
We note that the simplified tree shares some similarities with those found in QMDP \cite{thrunProbabilisticRobotics2002} and contingency planning \cite{HardyContingencyPlanning2013, AlsterdaContingency2019}.
Figure \ref{fig:simple_trajectory_tree} shows an example of the simplified trajectory tree with two branches. 
We can now discuss the formulation in more detail.

\subsection{Detailed Formulation} \label{subsec:details}
\subsubsection{Branches}
Our behavior planner generates multiple multi-vehicle trajectories associated with distinct equilibrium strategies.
Given that the interacting vehicle (IV) might be uncertain in terms of equilibrium strategies, we employ the branch MPC formulation to accommodate all potential equilibrium strategies.
In this formulation, each tree branch represents an equilibrium strategy, such as a Nash or Stackelberg strategy.

\subsubsection{Cost function} 
The motion planner tracks the reference trajectories generated by the behavior planner while maximizing the comfort level. 
Thus, we use the following stage cost:
\begin{align*}
    \ell &:= \ell_{\text{track}} + \ell_{\text{com}}, \\
    \ell_{\text{track}}(x_i^{\text{EV}}, u_i^{\text{EV}}) &:= \left\Vert x_i^{\text{EV}} - x_i^{\text{EV, ref}}  \right\Vert_Q^2 + \left\Vert u_i^{\text{EV}} - u_i^{\text{EV, ref}} \right\Vert_R^2,\\
    \ell_{\text{com}}(u_i^{\text{EV}}, u_{\text{p}(i)}^{\text{EV}}) &:= 
    \left\Vert u_i^{\text{EV}} - u_{\text{p}(i)}^{\text{EV}} \right\Vert_{R_{\text{com}}}^2,\; \forall i \in \mathcal{N}, 
\end{align*}
where $R_{\text{com}} \succ 0$, $Q \succcurlyeq 0$, and $R \succ 0$ represent weight matrices for comfort, state, and control input, respectively; 
we define $u_{\text{p}(0)}^{\text{EV}}$ as the executed control command from the last planning cycle. 
The penalty term on the change rate of the control input plays an important role in enhancing driving comfort. 
By penalizing $\left\Vert u_0^{\text{EV}} - u_{\text{p}(0)}^{\text{EV}} \right\Vert$, we reduce the executed control variation between two planning cycles. 
\subsubsection{State and input constraints} To ensure physical feasibility, we impose bounds on the state and control input:
\begin{equation*}
   x_{\text{lb}} \leq x_i^{\text{EV}} \leq x_{\text{ub}},\; \forall i \in \mathcal{N},\; u_{\text{lb}} \leq u_i^{\text{EV}} \leq u_{\text{ub}},\; \forall i \in \mathcal{N} \backslash \mathcal{L}.
\end{equation*}


\subsubsection{Collision avoidance} 
The footprint of a vehicle can be naturally represented by a rectangle.
Nevertheless, one challenge associated with this representation is the complexity of deriving a closed-form signed distance. Moreover, the gradient of the signed distance function is discontinuous, posing a challenge to the optimization solver. 
To circumvent these difficulties,
we use an overapproximation of
the shape of a vehicle that comprises the union of a collection of linked circles \cite{WangInteraction2023}.
The smooth collision avoidance constraints \eqref{eq:tree_state_constr} can be formulated as follows:
\begin{equation}    
\begin{aligned} 
    h(\boldsymbol{x}) = 
    (r^{\text{EV}} + r^{\text{IV}})^2 - \left\Vert c_i(x^{\text{EV}}) - c_j(x^{\text{IV}}) \right\Vert_2^2 \leq 0, \\i \in [1, n_c^{\text{EV}}],\; j \in [1, n_c^{\text{IV}}],
\end{aligned}
\end{equation}
where $r^{\text{EV}}$ and $r^{\text{IV}}$ represent the disc radii, $c: \mathbb{R}^n \rightarrow \mathbb{R}^2$ is a function that computes the center of the disc, 
and $n_c^{\text{EV}}$, $n_c^{\text{IV}}$ denote the number of discs used for approximating the vehicle footprints.
In addition to the IV, we enforce the same dynamic collision avoidance constraints on the other vehicles using the simulated trajectories provided by the behavior planner.

\section{Numerical Simulations} 

\begin{figure}[!t]
    \centering
    \includegraphics[width=0.9\columnwidth]{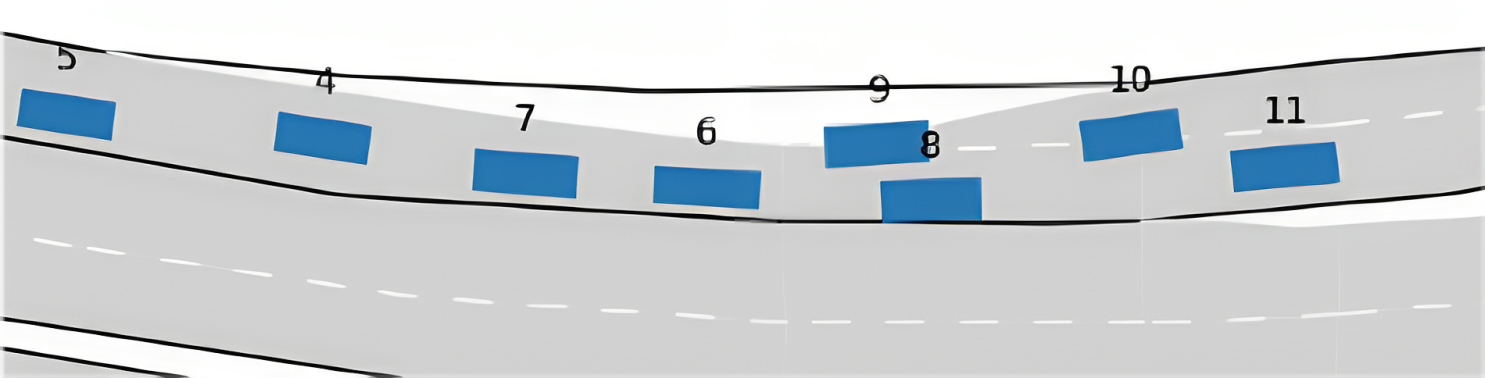}
    \caption{On-ramp merging scenario: vehicle 9 intends to reach the gap between vehicles 6 and 8. Since the gap is not sufficiently large, vehicle 9 has to identify the intention of vehicle 8 and make an appropriate decision.}
    \label{fig:on-ramp_merging}
    \vspace{-2.5mm}
\end{figure}

\label{sec:numerical_simulations}
\subsection{Benchmarking}
We validate our proposed game-theoretic planner using the INTERACTION dataset \cite{interactiondataset2019}. We consider the on-ramp merging scenario shown in Figure \ref{fig:on-ramp_merging}, where the ego vehicle must complete the lane change before reaching the end of the current lane. 
Considering various factors, such as traffic speeds, vehicle sizes, and behavior modes, we select 100 test scenarios from the dataset and mark the ego vehicle for each track.
Each test scenario on the dataset has a duration of $T_{\text{tr}} = \SI{4}{s}$ with a discretization step of $\Delta t = \SI{100}{ms}$, and the number of timestamps is $N_{\text{ts}} = T_{\text{tr}} / \Delta t$.
We control the ego vehicle using our proposed planner, while the motion of surrounding vehicles is simulated either by replaying trajectory data (nonreactive mode) or using the IDM (reactive mode).
We compare the proposed GT-BMPC with five baseline methods:
\begin{itemize}
    \item NE-MPC \cite{LuyaoMaxtrixGame2023}, combining a Nash equilibrium-seeking behavior planner with a traditional MPC motion planner \cite{HowellALTRO2019}. 
    This method does not consider the strategy misalignment. 
    \item SE-MPC \cite{zhangGameTheoreticModel2020, weiGameTheoreticMerging2022}, which searches for the Stackelberg equilibrium in the behavior planner and then uses a traditional MPC for motion planning.
    \item Y-MPC, which comprises a behavior planner and a traditional MPC motion planner. The behavior planner assumes that the vehicle group always decides to \texttt{yield}.
    \item PGame \cite{liu2023potential}, a finite potential game framework, using sampling-based trajectories \cite{werling2012optimal} as actions.
    \item iLQGame \cite{FridovichILQG2020}, which jointly plans Nash equilibrium trajectories for all vehicles.
\end{itemize}

The quality of the trajectories is evaluated using the following metrics:
\begin{enumerate*}[label=(\roman*)]
    \item We assess the safety level via collision rate and Time-to-Collision $(TTC)$ \cite{LiLevelKGame2022, Nuplan2022}. 
    The collision rate is defined as the proportion of the number of cases, in which a collision happens, to the total number of cases. 
    $TTC$ refers to the time it takes for two vehicles to collide if they maintain their current speed and heading. 
    Note that, akin to those in \cite{Nuplan2022} and \cite{JiaoTTC2023}, $TTC$ utilized here accounts for both longitudinal and lateral directions.
   
    \item 
    Regarding progress, we project the vehicle position at the final timestamp $p_{N_{\text{ts}}}$ onto the target lane to determine the lateral distances $d_{N_{\text{ts}}}$.
    A smaller lateral distance indicates better lateral progress.

    \item We employ the root mean squared absolute jerk, maximum absolute jerk, and root mean squared angular acceleration as metrics to measure comfort level.
    With the velocity profile of a track, we numerically approximate jerk at one timestamp as $\left\vert (v_{k-1}^{\text{EV}} - 2 v_{k}^{\text{EV}} + v_{k+1}^{\text{EV}} ) / \Delta t^2 \right\vert$.
    Similarly, the approximated angular acceleration is calculated by $\left\vert (\theta_{k-1}^{\text{EV}} - 2 \theta_{k}^{\text{EV}} + \theta_{k+1}^{\text{EV}} ) / \Delta t^2 \right\vert$, where $\theta$ is the heading angle.
    \item We use average displacement error (ADE) to measure the distance between the generated trajectory and the ground truth \cite{dingEPSILONEfficientPlanning2022}. Specifically, $\text{ADE}:= \sum_{k = 1}^{N_{\text{ts}}} \left\Vert p_k^{\text{EV}} - p_k^{\text{EV, GT}} \right\Vert / N_{\text{ts}}$, where $p_k^{\text{EV, GT}}$ denotes the ground truth position of the ego vehicle.
  
\end{enumerate*}

\subsection{Implementation details} \label{subsec:implementation_details}
\subsubsection{Game-theoretic behavior planner}
We use a planning horizon of $T_{\text{bp}}=25$, a discretization step of $\Delta t_{\text{bp}} = \SI{0.2}{s}$, a decision time period of $\Delta h = \SI{1}{s}$ and a decision horizon of $H=5$. 
In other words, the behavior planner looks ahead for $\SI{5}{s}$ and makes a semantic-level decision every $\SI{1}{s}$. We run the behavior planner at \SI{5}{\hertz} in a receding horizon fashion. 

We select the vehicle nearest to the ego vehicle as the target vehicle (SV1). 
The vehicles ahead of and behind the target vehicle are denoted as SV0 and SV2, respectively.
The gap in front of the target vehicle is denoted as $\texttt{Gap1}$, while the gap behind is referred to as $\texttt{Gap2}$, as shown in Figure \ref{fig:semantic_actions}.
We label the current lane where the ego vehicle stays as $\texttt{Gap0}$. 
As discussed in Section \ref{subsec:scenario_generation}, there are three desired lateral positions: two center lines and one probing line.

We consider the vehicle ahead of the ego vehicle as a dynamic obstacle, indicating that the ego vehicle intends to maintain a safe distance from it.
\textcolor{black}{The surrounding vehicles, SV0, SV1, and SV2 are treated as a vehicle group. 
The potential interacting vehicles in this group are the SV1 and SV2.
We assume that the interacting vehicle has two longitudinal actions: $\texttt{Assert}$ and $\texttt{Yield}$.
To represent the actions, we supply distinct parameters to the modified IDM, which controls the motion of the surrounding vehicle.
For instance, to model a yielding vehicle, in addition to the willingness indicator $\beta$ in \eqref{eq:idm_distance}, we set the minimum spacing and desired time headway to relatively large values.
}


\subsubsection{BMPC}
The motion planner operates at \SI{10}{\hertz} with a discretization step of \SI{0.1}{s} and a planning horizon of \SI{4}{s}. 
The reference trajectories for the ego vehicle, along with the probability associated with each branch, are provided by the behavior planner. 
To exploit the sparse structure of the optimal control problem, we develop a tailored BMPC solver by extending the iLQR solver \cite{TassaiLQR2012} to its trajectory tree version, known as iLQR-tree \cite{DaComprehensive2022, LiMultipolicy2023}. 
We handle state and input constraints using the augmented Lagrangian relaxation \cite{HowellALTRO2019}. 

\subsubsection{Computational performance}
All simulations are conducted on a laptop with a \SI{2.30}{\giga\hertz} Intel Core i7-11800H processor and \SI{16}{\giga\byte} RAM.
We implement both behavior planner and motion planner in C++.
The behavior planner requires approximately \SI{10.5}{\milli\second} to generate the multi-vehicle trajectories. 
The real-time performance can be attributed to two factors:
\begin{enumerate*}[label=(\roman*)]
    \item Sampling the semantic-level action space, rather than the control input space, resulting in fewer samples. 
    \item Parallelizing the process of forward simulation using the OpenMP API. 
\end{enumerate*}
Regarding the motion planner, the average computation time for the BMPC problem is \SI{1.11}{ms}, while the peak computation time reaches \SI{7.23}{ms}.
We further enhance the computational efficiency by computing the backward sweeps from each leaf node to the shared node in parallel via OpenMP API.
As a result, we achieve a reduction of \SI{9.9}{\percent} in the mean computation time (\SI{1.00}{ms}) and a decrease of \SI{19.2}{\percent} in the maximum computation time (\SI{5.84}{ms}).



\begin{figure}[!t]\captionsetup[subfigure]{font=scriptsize}
\centering
\subfloat[]{\includegraphics[width=0.45\columnwidth]{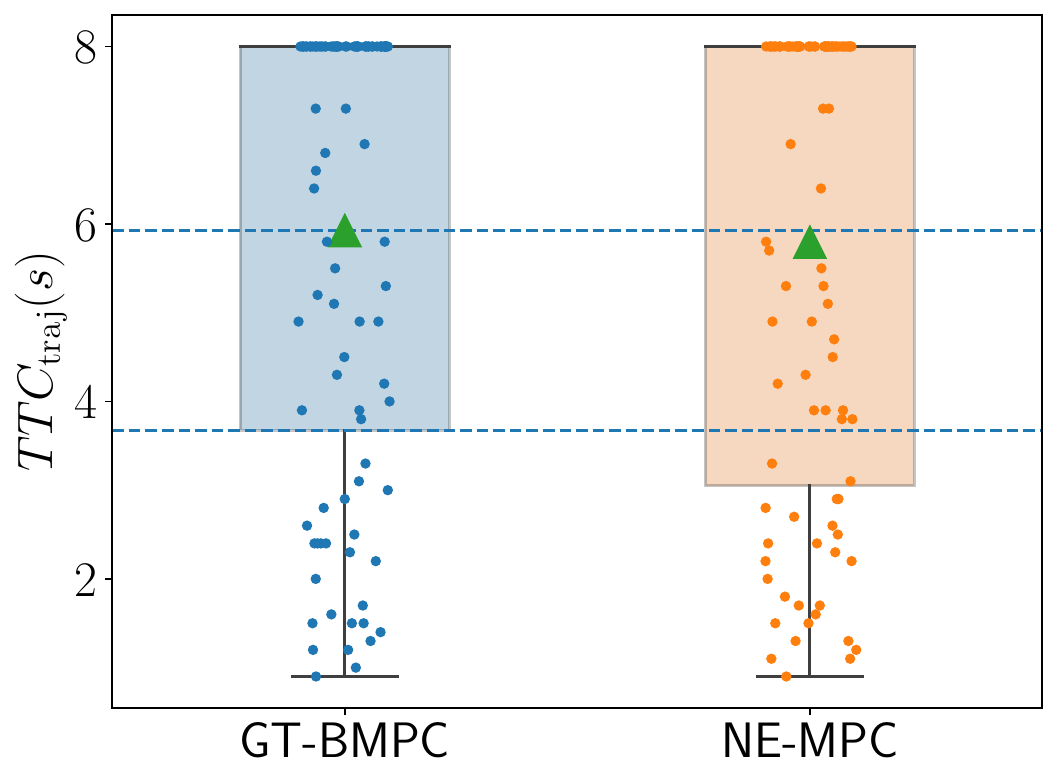} \label{fig:ttc}}
\subfloat[]{\includegraphics[width=0.47\columnwidth]{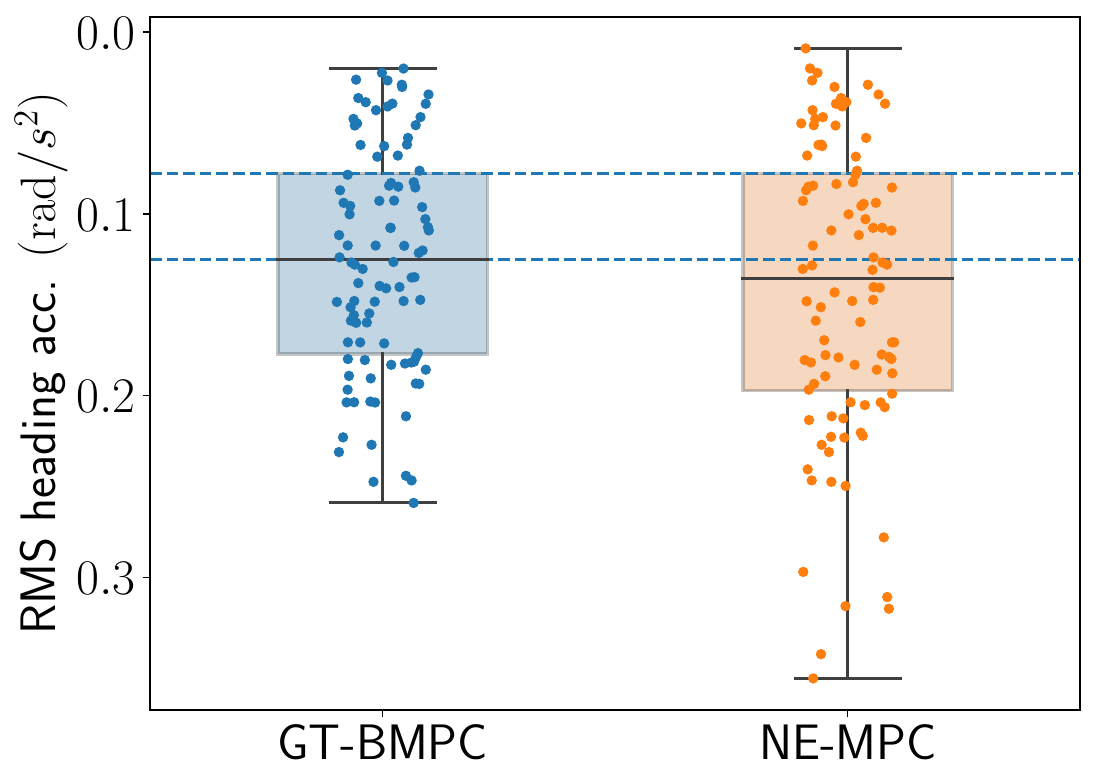} \label{fig:heading_acc}}

\caption{(a): Box plot of $TTC_{\text{traj}}$.
We limit $TTC_{\text{traj}}$ to \SI{8}{\second} for visualization. 
$TTC_{\text{traj}}$ represents the minimum $TTC$ values among all timestamps of one pair of trajectories. 
To obtain it, we first calculate $TTC_k$ at timestamp $k$. 
In a lane-merging scenario shown in Fig. \ref{fig:on-ramp_merging}, the EV (vehicle 9) selects the gap between the leading vehicle (LV, vehicle 6) and the following vehicle (FV, vehicle 8).  
To consider both surrounding vehicles, we calculate two $TTC$ values at each timestamp: one between the EV and LV as $TTC_{\text{EV-LV},k}$, and another between the FV and EV as $TTC_{\text{FV-EV},k}$. 
Next, we choose $TTC_{k} = \min \{ TTC_{\text{EV-LV},k}, TTC_{\text{FV-EV},k}\}$.
After calculating $TTC_k$ at all timestamps, we select the minimum from $\{TTC_k\}$ as $TTC_{\text{traj}}$.
The triangle marker represents the mean value. (b): Box plot of root mean squared heading acceleration.}
\vspace{-2.5mm}
\end{figure}

\subsection{Statistical results}
We compare the proposed planner with the baseline planners across 100 test scenarios. Table \ref{tb:statistical_results} shows the mean values of the considered metrics for these scenarios. 
\subsubsection{Nonreactive simulation}
Both GT-BMPC and NE-MPC operate safely across all test scenarios due to the consideration of diverse driving behaviors, whereas the others do not.
Since Y-MPC, SE-MPC, and PGame tend to assume that the interacting vehicle behaves cooperatively, collisions are likely to occur if the vehicle fails to yield.
Moreover, we compute $TTC_{\text{traj}}$ to evaluate the safety level further, and the statistical results are illustrated in Figure \ref{fig:ttc}. 
The mean and lower quartile of GT-BMPC exceed those of NE-MPC, indicating that GT-BMPC generates trajectories with a larger safety margin across the majority of test scenarios.
GT-BMPC shows slightly reduced lateral progress because it often plans a probing trajectory to account for the various potential behaviors of surrounding vehicles in the beginning. Such a motion plan can enhance safety in certain situations (see Section \ref{subsec:case_study}).
Regarding comfort assessment, all the planners, except for PGame and iLQGame, achieve comparable performance. 
Figure \ref{fig:heading_acc} illustrates that GT-BMPC outperforms NE-MPC with respect to heading acceleration. 
This improvement is attributed to the ability of the trajectory tree to mitigate the issue arising from abrupt changes in semantic-level actions between two planning cycles. 
Finally, the low ADE values achieved by GT-BMPC, NE-MPC, and SE-MPC suggest that the ego vehicle controlled by these planners exhibits more human-like driving behavior.
To sum up, our proposed planner balances various metrics and plans safer trajectories, especially when the behavior modes of surrounding vehicles are time-varying, which will be further discussed in Section \ref{subsec:case_study}. 

\subsubsection{Reactive simulation}
Collision rates decrease significantly due to the reactive behavior of surrounding vehicles, and all planners achieve improved lateral progress. 
Additionally, Y-MPC and SE-MPC perform well in reactive simulations, as their assumptions about the cooperative behavior of surrounding vehicles are satisfied.

\begin{figure*}[!t]\captionsetup[subfigure]{font=scriptsize}
\centering
\subfloat[GT-BMPC]{\includegraphics[width=0.30\textwidth]{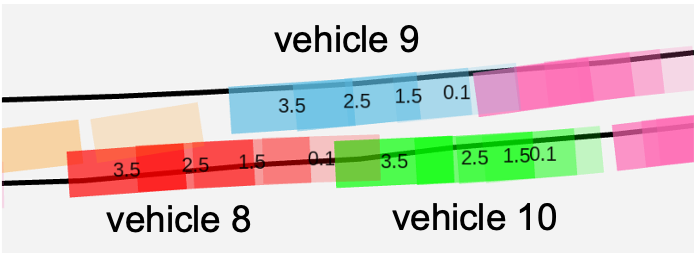}%
\label{fig:GTBP-BMPC}
}
\hfill
\subfloat[NE-MPC]{\includegraphics[width=0.3\textwidth]{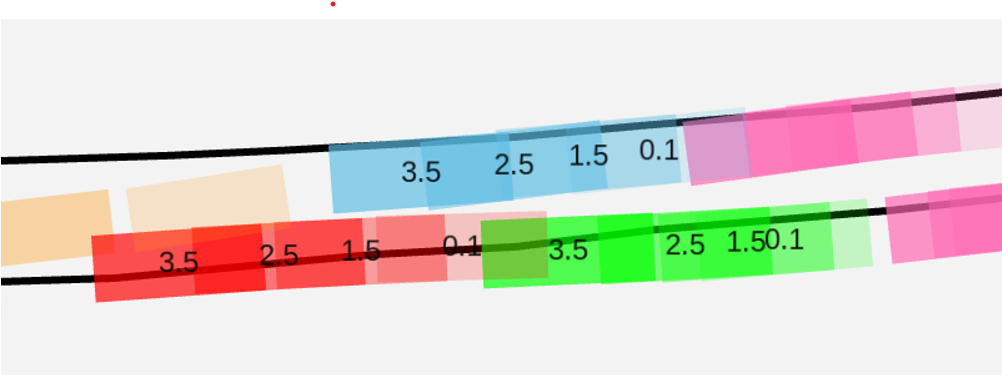}%
\label{fig:GTBP-MPC}
}
\hfill
\subfloat[Y-MPC]{\includegraphics[width=0.3\textwidth]{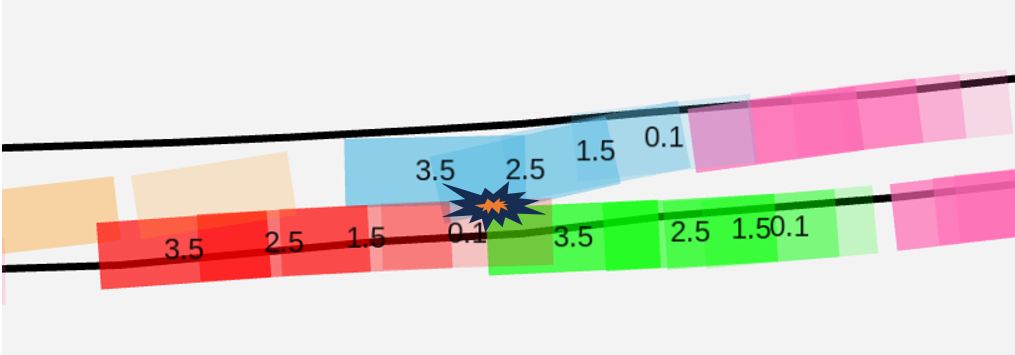}%
\label{fig:YBP-MPC}
}
\vfill
\subfloat[Heading]{\includegraphics[width=0.30\textwidth]{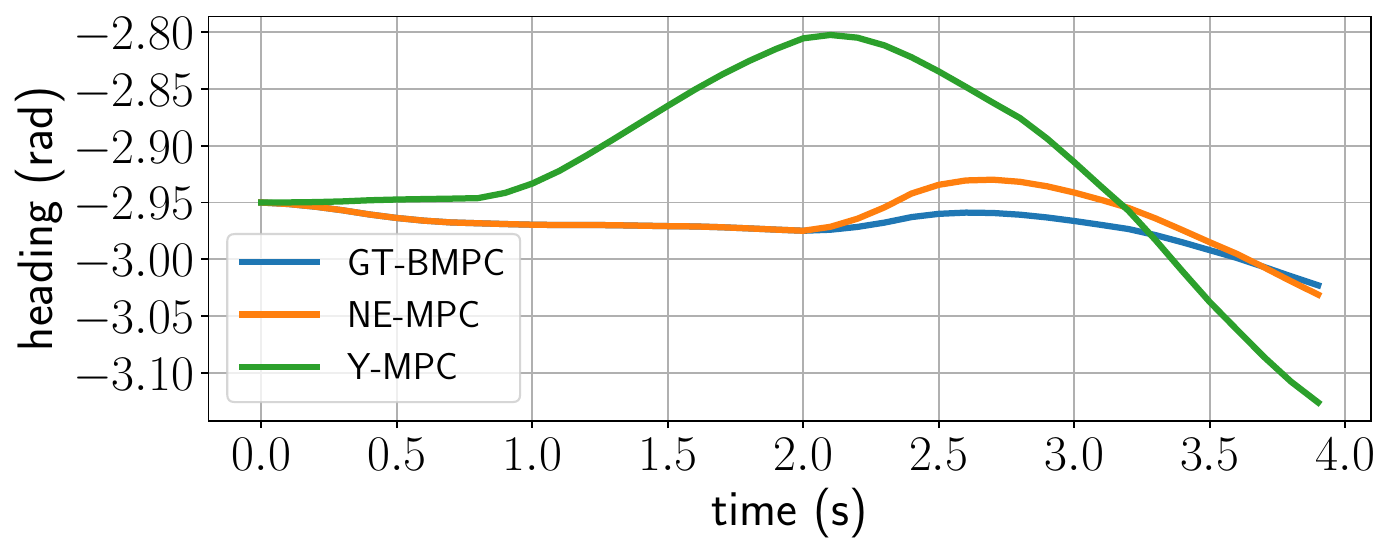}}%
\hfill
\subfloat[Acceleration]{\includegraphics[width=0.30\textwidth]{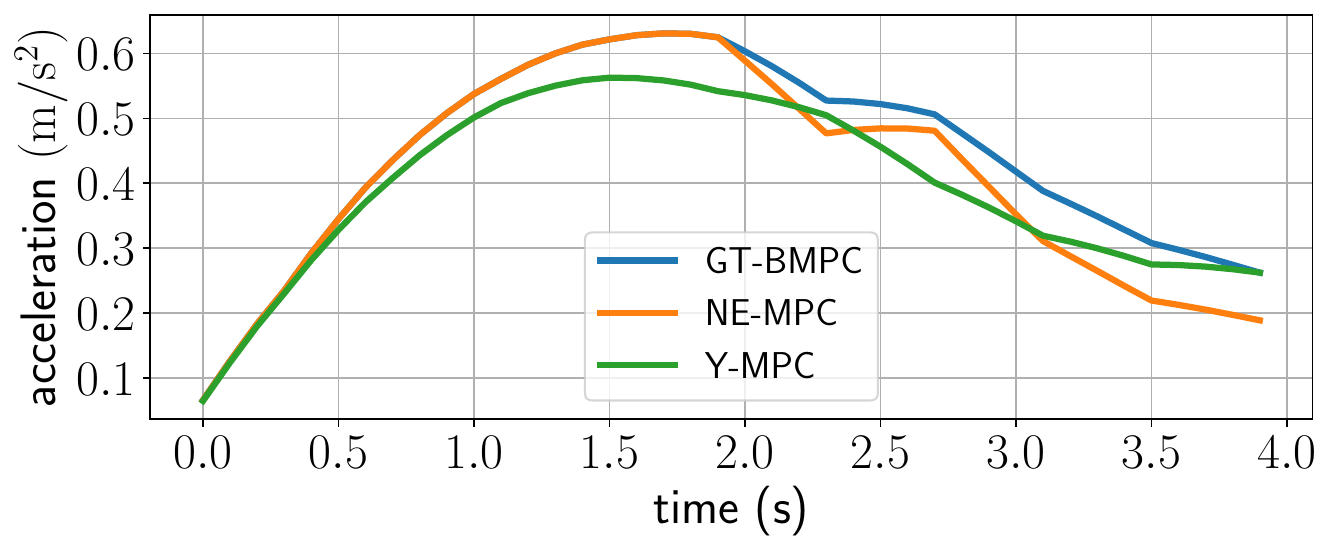}}%
\hfill
\subfloat[Steering angle]{\includegraphics[width=0.30\textwidth]{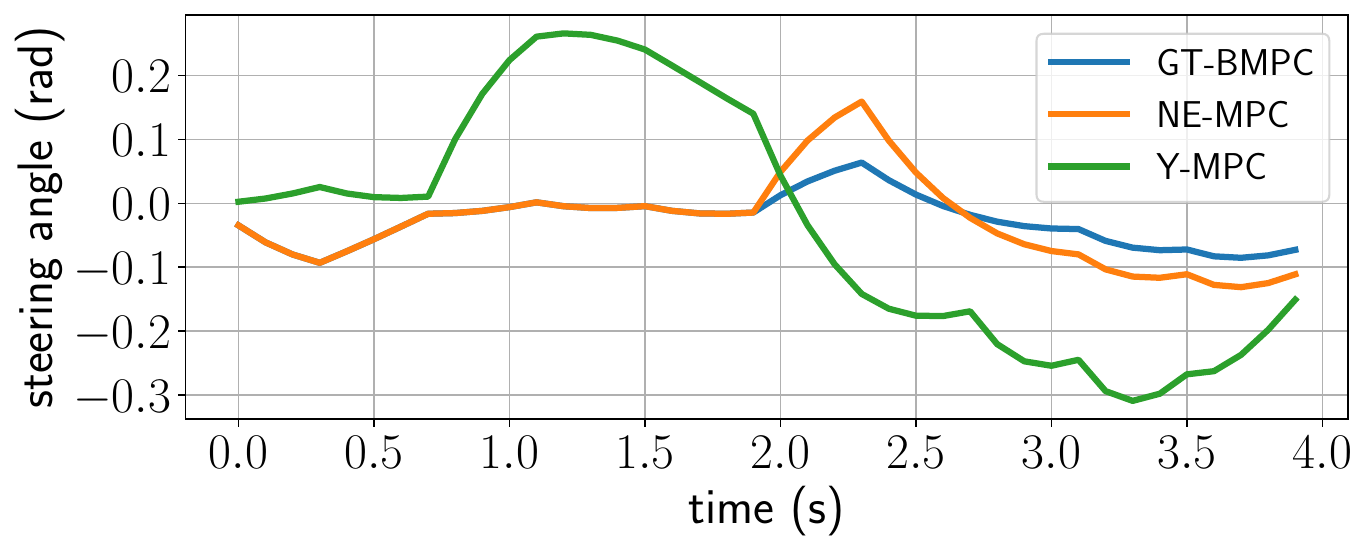}}%

\caption{
Simulation results of the selected scenario.
(a)-(c): Closed-loop trajectories. 
The ego vehicle is highlighted in blue, while potential interacting vehicles are marked in red and green.
The orange vehicle is treated as a dynamic obstacle, and non-interacting vehicles are marked in pink.
The number on each vehicle represents a simulation timestamp in seconds.
(d)-(f): Motion profiles of the ego vehicle.
The abrupt changes in the plots are likely attributed to replanning.
The control commands generated by GT-BMPC are smoother, leading to a more comfortable trajectory. 
}
\label{fig:sc168}
\end{figure*}

\setlength{\tabcolsep}{1.5pt}
\begin{table*}[ht]
  \centering
  \caption{Statistical Results}
  \label{tb:statistical_results}
  \begin{tabular}{@{}lcccccccccccc@{}}
    \toprule
    \multicolumn{1}{c}{\multirow{2}{*}{Metric}} & \multicolumn{6}{c}{Nonreactive} & \multicolumn{6}{c}{Reactive} \\ 
    \cmidrule(lr){2-7} \cmidrule(lr){8-13}
    & GT-BMPC & NE-MPC & SE-MPC & Y-MPC & PGame & iLQGame & GT-BMPC & NE-MPC & SE-MPC & Y-MPC & PGame & iLQGame \\ \midrule
    Collision Rate (\unit{\percent}) & \cellcolor{gray!15}0 & \cellcolor{gray!15}0 & 1 & 1 & 6 & 5 & \cellcolor{gray!15}0 & \cellcolor{gray!15}0 & \cellcolor{gray!15}0 & \cellcolor{gray!15}0 & 2 & 1 \\
    Lateral Progress (\unit{\meter}) & 1.21 & 1.20 & 1.13 & 1.13 & \cellcolor{gray!15}0.78 & 0.92 & 1.09 & 1.07 & 0.97 & 0.96 & \cellcolor{gray!15}0.55 & 0.60 \\
    RMS Abs. Jerk (\unit{\meter/\cubic\second}) & 0.21 & 0.21 & 0.21 & \cellcolor{gray!15}0.20 & 0.42 & 1.45 & \cellcolor{gray!15}0.24 & \cellcolor{gray!15}0.24 & \cellcolor{gray!15}0.24 & \cellcolor{gray!15}0.24 & 0.39 & 1.12 \\
    Max Abs. Jerk (\unit{\meter/\cubic\second}) & 0.52 & 0.52 & 0.49 & \cellcolor{gray!15}0.48 & 5.77 & 2.97 & 0.60 & 0.60 & 0.57 & \cellcolor{gray!15}0.56 & 4.74 & 2.98 \\
    RMS Heading Acc. (\unit{\radian/\square\second}) & \cellcolor{gray!15}0.12 & 0.14 & 0.13 & \cellcolor{gray!15}0.12 & 0.14 & 0.21 & 0.15 & 0.17 & 0.15 & 0.15 & \cellcolor{gray!15}0.14 & 0.16 \\
    ADE (\unit{\meter}) & 0.71 & \cellcolor{gray!15}0.70 & 0.72 & 0.75 & 0.91 & 1.19 & N/A & N/A & N/A & N/A & N/A & N/A \\
    \bottomrule
  \end{tabular}
  \vspace{-2mm}
\end{table*}

\begin{table}
  \begin{center}
    \caption{Case Study}
    \label{table:case_study}
    \begin{tabular}{@{}cccc@{}}
    \toprule
    Metric & GT-BMPC & NE-MPC & Y-MPC \\ \midrule
    \begin{tabular}[c]{@{}c@{}} Collision-free \\ $TTC_{\text{traj}}$ (\unit{\second}) \\ RMS Abs. Jerk (\unit{\meter/\cubic\second}) \\ Max Abs. Jerk (\unit{\meter/\cubic\second}) \\ RMS Heading Acc. (\unit{\radian/\square\second})  \\
    \end{tabular} & 
    \begin{tabular}[c]{@{}c@{}} \checkmark \\ \cellcolor{gray!15}4.10  \\ \cellcolor{gray!15}0.23 \\ \cellcolor{gray!15}0.61 \\ \cellcolor{gray!15}0.09 \end{tabular} & 
    \begin{tabular}[c]{@{}c@{}} \checkmark \\ 2.50 \\ 0.25 \\ \cellcolor{gray!15}0.61 \\ 0.18 \end{tabular}  & 
    \begin{tabular}[c]{@{}c@{}} \xmark \\ N/A \\ N/A\\  N/A\\ N/A\\ \end{tabular}    \\
    \bottomrule
    \end{tabular}
  \end{center}
  \vspace{-3mm}
\end{table}

\subsection{Case study} \label{subsec:case_study}
We choose one interesting scenario to demonstrate the effectiveness of the integration of the game-theoretic behavior planner and the BMPC framework. 
The simulation results are illustrated in Figure \ref{fig:sc168}.
The ego vehicle (vehicle 9) faces two major challenges in this scenario. First, the feasible driving space is limited due to the narrow environment. Secondly, the actual behavior mode of vehicle 10 is variable. 
Specifically, vehicle 10 decelerates during the initial \SI{1}{\second} but speeds up afterward. 
Consequently, all the ego vehicles in Figure \ref{fig:sc168} decide to change the lane in the beginning but finally cancel the lane change. 
In Figure \ref{fig:YBP-MPC}, 
as the ego vehicle does not realize that merging is dangerous in time, a collision with vehicle 10 becomes unavoidable. 
On the contrary, the closed-loop trajectories generated by GT-BMPC and NE-MPC are collision-free due to the consideration of multi-modal behaviors. 
We observe that the distance between the ego vehicle and vehicle 10 at \SI{3.5}{\second} in Figure \ref{fig:GTBP-BMPC} is larger than that in Figure \ref{fig:GTBP-MPC}, and the corresponding $TTC_{\text{traj}}$ values in Table \ref{table:case_study} further affirm that GT-BMPC outputs a safer trajectory, as it allows the ego vehicle to safely return to the original lane by adhering to another equilibrium strategy.
Furthermore, GT-BMPC outperforms NE-MPC in terms of RMS absolute jerk and RMS heading acceleration. 


\section{Conclusion} \label{sec:conclusion}
In this paper, we present a game-theoretic planning framework to address lane-merging problems in interactive environments.
Our approach explicitly models the interaction between vehicles using a matrix game and tackles the issue caused by the uncertain behaviors of the surrounding vehicles via BMPC.
Our validation study on the INTERACTION dataset indicates the necessity of interaction modeling and demonstrates the superior performance of our proposed method compared to the state-of-the-art approaches.
In future work, we aim to enhance the computational efficiency and robustness of the BMPC planner. 
We plan to explore the recursive feasibility of the BMPC planner under certain technical assumptions, such as the existence of controlled invariant sets and the ability to gather information incrementally. 
Additionally, we intend to extend this planning framework to handle intersection-crossing scenarios. Furthermore, we plan to validate the entire planning framework on a hardware platform.


\printbibliography

\end{document}